\documentclass[aps, prd, amsmath, floats, floatfix, twocolumn, nofootinbib, superscriptaddress]{revtex4-1}
\usepackage[pdftex]{graphicx}
\usepackage{color}
\usepackage{latexsym}
\usepackage[colorlinks]{hyperref}
\usepackage{amsthm,amsmath,amssymb}
\usepackage{mathrsfs}
\usepackage{makecell}

\newcommand{\be}{\begin{eqnarray}}
\newcommand{\ee}{\end{eqnarray}}

\newcommand{\aj}{Astron.\ J.\ }
\newcommand{\apjl}{Astrophys.\ J.\ }

\newcommand{\aap}{Astron.\ Astrophys.\ }
\newcommand{\araa}{Annu.\ Rev.\ Astron. Astrophys.\ }

\newcommand{\mnras}{Mon.\ Not.\ R.\ Astron.\  Soc.\ }
\newcommand{\pasj}{Publ.\ Astron.\ Soc.\ Japan\ }
\newcommand{\jcap}{JCAP~}
\newcommand{\ssr}{Space Sci.\ Rev.\ }
\newcommand{\iaucirc}{IAU Circ.}

\begin{document}

\title{Impact of reflection Comptonization on X-ray reflection spectroscopy:\\the case of EXO~1846--031}

\author{Songcheng~Li}
\affiliation{Center for Astronomy and Astrophysics, Center for Field Theory and Particle Physics, and Department of Physics,\\
Fudan University, Shanghai 200438, China}

\author{Honghui~Liu}
\affiliation{Center for Astronomy and Astrophysics, Center for Field Theory and Particle Physics, and Department of Physics,\\
Fudan University, Shanghai 200438, China}

\author{Cosimo~Bambi}
\email[Corresponding author: ]{bambi@fudan.edu.cn}
\affiliation{Center for Astronomy and Astrophysics, Center for Field Theory and Particle Physics, and Department of Physics,\\
Fudan University, Shanghai 200438, China}
\affiliation{School of Natural Sciences and Humanities, New Uzbekistan University, Tashkent 100007, Uzbekistan}

\author{James~F.~Steiner}
\affiliation{Center for Astrophysics, Harvard \& Smithsonian, Cambridge, MA 02138, United States}

\author{Zuobin~Zhang}
\affiliation{Center for Astronomy and Astrophysics, Center for Field Theory and Particle Physics, and Department of Physics,\\
Fudan University, Shanghai 200438, China}

\begin{abstract}
Within the disk-corona model, it is natural to expect that a fraction of reflection photons from the disk are Compton scattered by the hot corona (reflection Comptonization), even if this effect is usually ignored in X-ray reflection spectroscopy studies. We study this effect by using \textit{NICER} and \textit{NuSTAR} data of the Galactic black hole EXO~1846--031 in the hard-intermediate state with the model {\tt simplcutx}. Our analysis suggests that a scattered fraction of order 10\% is required to fit the data, but the inclusion of reflection Comptonization does not change appreciably the measurements of key-parameters like the black hole spin and the inclination angle of the disk. 
\end{abstract}

\maketitle


\section{Introduction}

\begin{figure}
	\centering
	\includegraphics[width=0.95\linewidth]{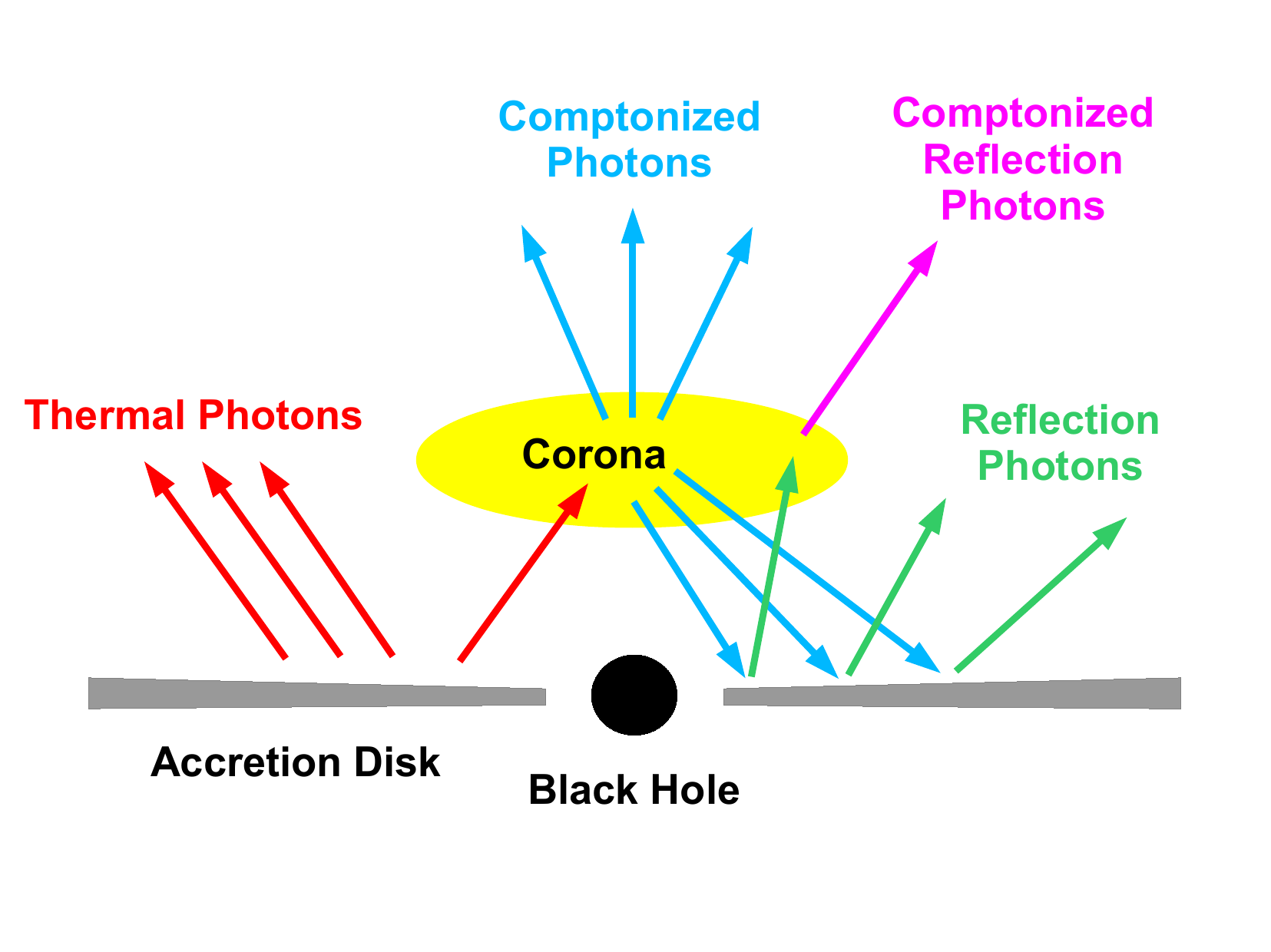}
	\vspace{-0.8cm}
	\caption{Disk-corona system with reflection Comptonization.}
	\label{corona0}
\end{figure}

In the disk-corona model, a black hole accretes from a geometrically thin and optically thick accretion disk, as shown in Fig.~\ref{corona0}~\cite{X-ray_spectroscopy, Fabian_1995_MNRAS_277L, Zoghbi_2010_MNRAS_401, Risaliti_2013_Natur_494}. The disk can be described by the Novikov-Thorne model~\cite{N-T_model_1973} and has a multicolor blackbody spectrum. A portion of the thermal photons are subsequently scattered by hot electrons residing in the corona, yielding a power-law component in the spectrum. The power-law component illuminates the disk, thereby producing a reflection component marked by fluorescent emission lines in the soft X-ray band and a Compton hump with a peak at 20-30~keV~\cite{2005MNRAS.358..211R,2010ApJ...718..695G}. With high quality data and the correct astrophysical model, X-ray reflection spectroscopy can be a powerful tool to study the morphology of the accreting system and the spacetime around the black hole~\citep{X-ray_spectroscopy}.

The corona plays a vital role in the process of spectrum generation within this paradigm. In the Comptonization of photons emitted from the accretion disk, we often consider Comptonization of thermal photons from the disk, ignoring the reflection photons. Reflection photons should also undergo Compton scattering off electrons in the corona above the accretion disk~\cite{2016ApJ...829L..22S,simplcut}, as shown in Fig.~\ref{corona0} by the magenta arrow. Such a reflection Comptonization seems to be unavoidable, but it is normally overlooked in the spectral analysis of accreting black holes. \citet{Wilkins_2015_MNRAS_448} showed that an extended corona in active galactic nuclei (AGN) should scatter the reflection photons and a similar conclusion can be expected in the case of stellar-mass black holes in X-ray binary systems. The fraction of reflection photons that are Compton scattered by the corona should depend on the extension of the corona and its optical depth. If the corona covers a sufficiently large fraction of the inner part of the accretion disk and its optical depth is not low, the reflection Comptonization is expected to have an impact on the spectral analysis and on the estimates of the properties of the accreting system.

In this paper, we present a study of the impact of the reflection Comptonization on the analysis of a reflection dominated spectrum of a Galactic black hole, EXO~1846--031. We use {\tt simplcutx} \cite{simplcut} to describe any Compton scattering in the corona. We compare the measurements of the properties of the spectrum with and without including the effect of reflection Comptonization.

EXO~1846--031 is a black hole X-ray binary discovered by \textit{EXOSAT} on April 3, 1985 \citep{EXO1846_1985}. After a quiescent period of over two decades, it experienced a new outburst in 2019, which was first detected by \textit{MAXI}/GSC on July 23, 2019 \cite{EXO1846_2019_MAXI}. This outburst was also observed by other missions during the following few weeks, such as \textit{Swift} \cite{EXO1846_2019_Swift}, \textit{NICER} \citep{EXO1846_2019_NICER}, and \textit{NuSTAR} \cite{EXO1846_2019_NuSTAR}. During the hard to soft state transition \citep{EXO1846_2019_insight-HXMT_hard-to-soft}, the source showed a spectrum with strong reflection features \cite{EXO1846_2019_NuSTAR} and quasi-periodic oscillations (QPOs) \citep{EXO1846_2019_insight-HXMT_QPO}. \citet{Draghis_2020_07} found that the black hole in EXO~1846--031 has a spin parameter very close to 1 and the inclination angle of its accretion disk (i.e., the angle between the normal to the disk surface and our line of sight) is high, close to 70~deg. Thanks to the very strong reflection features in the \textsl{NuSTAR} spectrum, this source has been widely used to test new reflection models (see, e.g., Refs.~\cite{Abdikamalov_2021, Tripathi_2021_relxillNK}) as well as different spacetime metrics from modified gravity theories (see, e.g., Refs.~\cite{Tripathi_2021_gravity, Yu_2021, Gu_2022, Tao_2023}). \citet{Abdikamalov_2021} showed that a radial disk ionization profile can improve the fit of the spectrum around 7~keV.
QPOs were studied in \citet{IHEP_2021} and the possible existence of a disk wind was studied in \citet{Wang_2021}.

Past studies have shown that EXO~1846--031 exhibits strong reflection features. In the present manuscript, we still analyze the \textit{NuSTAR} spectrum of the 2019 outburst, but we also incorporate soft X-ray data from \textit{NICER} and perform joint fitting analysis to explore whether this will lead to any differences in the final results.

The content of the manuscript is as follows. In Section~\protect\ref{reduction}, we present the observation information and data reduction process. In Section~\ref{model}, we describe the two model configurations to study the impact of reflection Comptonization and present the spectral analysis results of the observation of EXO~1846--031. We discuss our results in Section~\ref{discussion} and present our conclusions in Section~\ref{conclusion}.

\section{Observations and data reduction}
\label{reduction}

\begin{figure}
	\centering
	\includegraphics[width=0.95\linewidth]{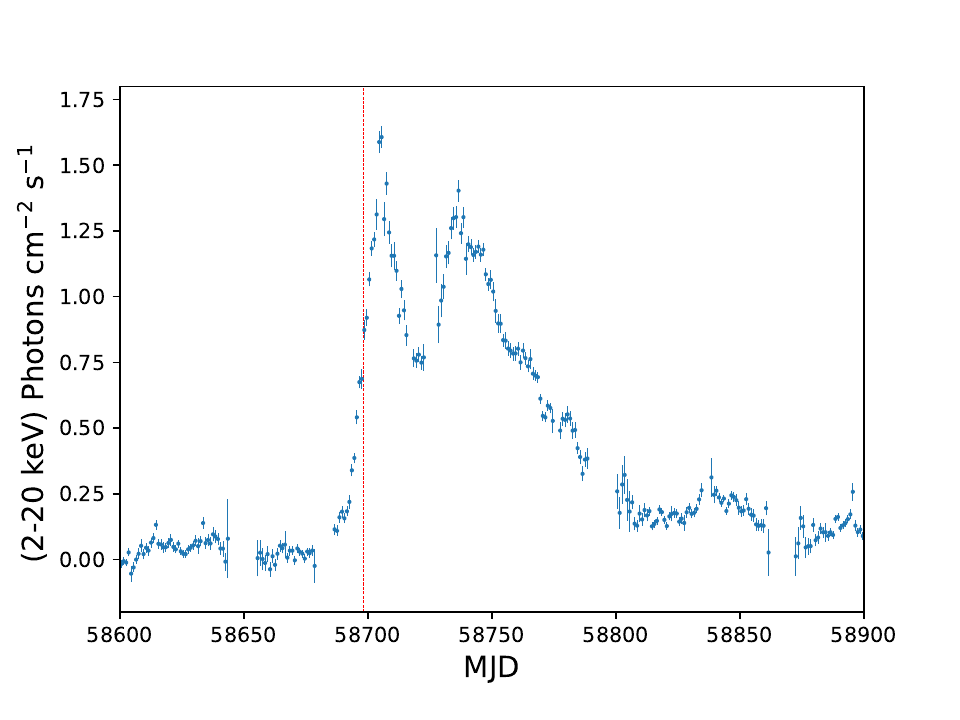}
	\vspace{-0.5cm}
	\caption{Light curve of EXO~1846--031 from Monitor of All-sky X-ray Image (MAXI). The red-dashed line marks the observation date of the data we used. Light curve data available from:\\ \url{http://maxi.riken.jp/star\_data/J1849-030/J1849-030.html}}
	\label{maxi_exo1846-031}
\end{figure}

EXO~1846--031 was observed by \textit{NuSTAR} and \textit{NICER} on August 3, 2019. The source was in a transitional phase from hard state to intermediate state \citep{Ren_2022}. The X-ray flux was about 46\% of the soft state peak (see Fig.~\ref{maxi_exo1846-031}). The details of the observations analyzed in our work are shown in Table~\ref{observation}.

\begin{table*}
	\centering
	\renewcommand\arraystretch{1.5}
	\caption{List of the observations analyzed in this paper.}\label{t-obs}
	\label{observation}
	\begin{tabular}{llccccc}
		\hline
		 & Source        & Mission & Observation ID & Observation Time    & Exposure (ks) & Count rate ($\rm s^{-1}$) \\
		\hline
		 & EXO~1846--031 & \textit{NuSTAR}  & 90501334002    & 2019-08-03 02:01:09 & 22.2          & 148.7             \\
		 &               & \textit{NICER}   & 2200760104     & 2019-08-03 02:08:58 & 0.911         & 376.1             \\
		\hline
	\end{tabular}\\
\end{table*}

For the \textit{NuSTAR} observation, we reduced the data with NuSTARDAS and the CALDB 20220301 \citep{NuSTAR_caliberation_2021}. We selected a circular region with a radius of $180^{\prime \prime}$ on both focal plane module A (FPMA) and focal plane module B (FPMB) detectors to extract the source spectra. In order to prevent influence from the source photons, a background region of comparable size was selected far from the source region. Afterwards, we used {\tt nuproducts} to generate the source and background spectra. We grouped the spectra with the optimal binning algorithm in \cite{opt-binning_2016} by using the {\tt ftgrouppha} task.

For the \textit{NICER} observation, the NICERDAS software suite was used to process the data. Extractions used standard screenings for the Sun angle, bright-Earth limb, and boresight. Any South Atlantic Anomaly (SAA) passages were excised.  Data from detectors 14, 34, and 54, were screened out, owing to intermittent calibration issues manifest from those detectors. All other of \textit{NICER}'s 49 additional working detectors were on, and from this set of events the detector ensemble distribution of X-ray events, undershoots, and overshoots were separately compared. There were no significant outliers, and so the remaining 49 active detectors were all used for spectral extraction. Undershoot levels were low for the observation in question, which ensures the robustness of the gain solution. The spectrum was adaptively binned to oversample the instrumental resolution by a factor of 2-3, and include at least 5 counts per bin. The 3C50 background model \citep{Remillard_2022_AJ} was used to produce background spectra.

\section{Data Analysis}
\label{model}

We use XSPEC v12.12.1 \citep{XSPEC} to analyze the spectra. We use two different datasets for the spectral fitting process: the spectra from \textit{NuSTAR} only (our ``\textit{NuSTAR} data''), and the spectra from both \textit{NuSTAR} and \textit{NICER} (our ``\textit{NuSTAR}+\textit{NICER} data'') for a joint fitting. The reason is to demonstrate the influence of soft X-ray data on the fitting.
The fits are made across the 3.0-79.0 keV energy band of \textit{NuSTAR} and the 1.0-10.0 keV energy band for \textit{NICER}. In the joint fitting of \textit{NuSTAR} and \textit{NICER} data, \textit{NuSTAR} data below 4.5 keV are ignored in order to avoid the mismatch between \textit{NICER} and \textit{NuSTAR} in the low energy band. Indeed, if we fit the data and we plot the residuals, we clearly see that the \textit{NICER} and \textit{NuSTAR} spectra are not consistent. Even in the study presented in Ref.~\cite{Nath_2024_ApJ_960_5N} on these \textit{NICER} and \textit{NuSTAR} spectra, the authors ignore the \textit{NuSTAR} data below 4~keV and a similar choice is adopted in Ref.~\cite{TaoLian_2019_ApJ_887_184T}, where the authors analyze \textit{NuSTAR} and \textit{Swift} data of GRS~1716--249 and ignore the \textit{NuSTAR} data below 4.5~keV. We do not know the exact origin of such a discrepancy in the low energy band; it may be some calibration issue in \textit{NuSTAR} \cite{Madsen_2020_arXiv, Chakraborty_2021_MNRAS_508}.

\begin{figure*}
	\centering
	\includegraphics[width=\linewidth]{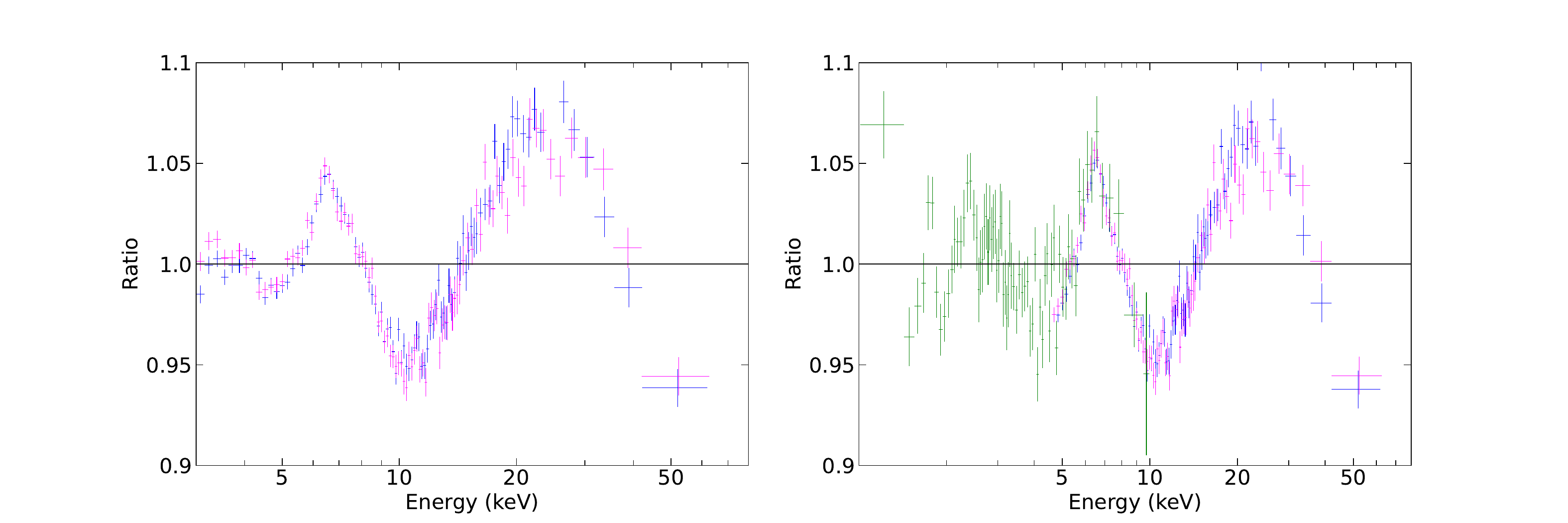}
	\caption{Data to best-fit model ratio plots for an absorbed disk blackbody and power-law model. The left panel is for the \textit{NuSTAR} data and the right panel is for the \textit{NuSTAR}+\textit{NICER} data. The blue and purple crosses are for the FPMA/\textit{NuSTAR} and FPMB/\textit{NuSTAR} data and the green crosses are for the \textit{NICER} data.
	\label{plot_ra:diskbb+nthcomp}}
\end{figure*}

First, we fit the data with an absorbed disk black-body and power-law to see the reflection features in the spectrum. In XSPEC language, the model reads 
\begin{equation}
	{\tt tbabs}*({\tt diskbb} + {\tt nthcomp}),
\end{equation}
where {\tt tbabs} describes the galactic absorption with the abundance table from \cite{tbabs_Wilms}, {\tt diskbb} \citep{pringle_1981, diskbb_1984} describes the thermal component, and {\tt nthcomp} \citep{nthcomp_zdziarski_1996, nthcomp_1999} describes the power-law component. We link $T_{\rm in}$ in {\tt diskbb} to $T_{\rm bb}$ in {\tt nthcomp}. We clearly see reflection features with a broad iron line around 7~keV and a Compton hump peaking around 20-30~keV (Fig.~\ref{plot_ra:diskbb+nthcomp}).

\subsection{Model 1: conventional modeling}

Previous study of the same observation shows that a reflection model that assumes a radial ionization gradient can fit the data better \citep{Abdikamalov_2021}. Therefore, we implement the {\tt relxillion\_nk} model that describes the ionization profile with a power-law:
\begin{equation}
	\xi(r) = \xi_{0}\left(\frac{R_{\rm in}}{r}\right)^{\alpha_{\xi}},
\end{equation}
where $\xi_0$ is the value of the ionization parameter at the inner edge of the disk, $R_{\rm in}$ is the radial coordinate of the inner edge of the disk, and $\alpha_\xi$ is the index of the power-law ionization profile.
The total model is {\tt tbabs*(diskbb+nthcomp+relxillionCp\_nk)} (Model~1), where {\tt relxillionCp\_nk} is a flavor of {\tt relxillion\_nk} that uses the {\tt nthcomp} Comptonization continuum for the incident spectrum.

\subsection{Model 2: self-consistent modeling with {\tt simplcutx}} 

Adopting a more self-consistent model structure, the power-law component is generated from Comptonization of the thermal component in the corona, and the reflection component originating from the disk surface will also be Comptonized~\citep{simplcut}.
In XSPEC language, the model expression is {\tt tbabs*simplcutx*(diskbb+relxillionCp\_nk)} (Model~2). We set the reflection fraction $R_{\rm F}$ in {\tt simplcutx} to 1 to avoid parameter degeneracy with the scattered fraction $f_{\rm SC}$\footnote{The scattering fraction $f_{\rm SC}$ is the fraction of thermal disk photons that are Compton-scattered by the corona into the power-law component. The reflection fraction $R_{\rm F}$ describes the flux of Compton-scattered photons directed back to the disk relative to the Compton flux that is transmitted to infinity. These two parameters are normally degenerate. See Ref.~\cite{simplcut} for more details.}.

\subsection{Analysis for the emissivity profile}

\begin{figure*}
	\centering
	\includegraphics[width=0.9\linewidth]{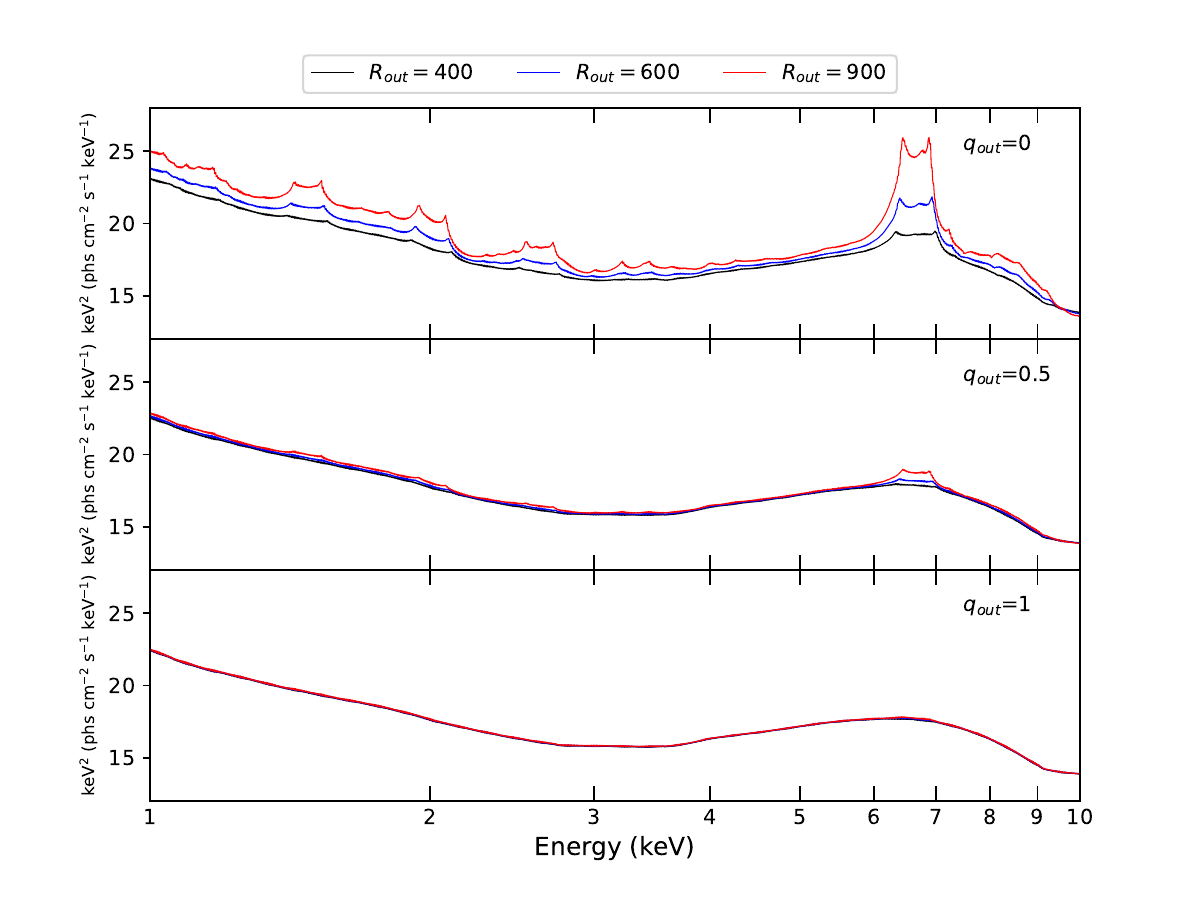}
	\vspace{-0.5cm}
	\caption{Relativistic reflection spectra with different values of $q_{\rm out}$ and $R_{\rm out}$. In all of these plots, the other parameters are set to the same specific values: $q_{\rm in} = 10$, $R_{\rm br}=6~R_{\rm g}$, $R_{\rm in}=R_{\rm ISCO}$, $a_{*}=0.998$, $i=75^\circ$, $\Gamma=2$, $E_{\rm cut}=100$~keV, $\log\xi=3.1$, $A_{\rm Fe}=1$.}\label{relxill_Rout_qout}
\end{figure*}

The emissivity profile depends on the geometry of the corona, which is unknown. Therefore, we model the emissivity profile using an empirical broken power-law ($\epsilon \propto 1/r^{q_{\rm in}}$ for $r < R_{\rm break}$, $\epsilon \propto 1/r^{q_{\rm out}}$ for $r > R_{\rm break}$). 
In previous studies on EXO~1846--031 \citep{Draghis_2020_07, Abdikamalov_2021}, the authors found that the \textit{NuSTAR} data can be fit well with a broken power-law emissivity profile with a high inner emissivity index ($q_{\rm in} \approx 10$) and an almost flat emissivity profile ($q_{\rm out} \approx 0$) at larger radii.
We note that the luminosity of an infinite disk is finite only if $q_{\rm out} > 2$.
Moreover, the theoretical reflection spectrum will obviously change with the variation of the disk's outer radius $R_{\rm out}$ if $q_{\rm out}$ is close to zero in reflection models like {\tt relxill\_nk}. This means that in our case with $q_{\rm out} \approx 0$ the reflection spectrum does depend on the exact value of the outer radial coordinate of the disk (see Fig.~\ref{relxill_Rout_qout} for an example).

Since the maximum value of the outer radius of the disk in our reflection model is 900 gravitational radii ($R_{\rm g}$) and if we leave it free it seems like the fit requires a larger value than this, we consider three different scenarios: $q_{\rm out}$ is free (case A), $q_{\rm out} = 3$ (case B), $q_{\rm out}$ is free and we add a distant reflector with {\tt xillverCp}, in which the ionization parameter is low ($\log\xi=0$; case C). 
We set $R_{\rm out}$ to its maximum value in the model in order to be sensitive even to the radiation emitted at relatively large radii.

All models we tested are summarized in Table~\ref{model_list}.

\begin{table*}
	\centering
	\renewcommand\arraystretch{1.5}
	\caption{Models used to analyze the \textit{NuSTAR} data of EXO~1846--031. As for the \textit{NuSTAR}+\textit{NICER} data, we just replace {\tt constant} with {\tt crabber*edge*xscat}.}
	\label{model_list}
	\vspace{0.3cm}
	\begin{tabular}{lcc}
		\hline
		model & \hspace{0.3cm} $q_{\rm out}$ \hspace{0.3cm} & XSPEC model for \textit{NuSTAR} data                                                                \\
		\hline
		1A    & free                & \tt{constant*tbabs*(diskbb + nthcomp + relxillionCp\_nk) }               \\
		1B    & $3^{*}$             & \tt{constant*tbabs*(diskbb + nthcomp + relxillionCp\_nk)}                 \\
		1C    & free                & \tt{constant*tbabs*(diskbb + nthcomp + relxillionCp\_nk + xillverCp)}     \\
		2A    & free                & \tt{constant*tbabs*simplcutx*(diskbb + relxillionCp\_nk)}               \\
		2B    & $3^{*}$             & \tt{constant*tbabs*simplcutx*(diskbb + relxillionCp\_nk)}               \\
		2C    & free                & \tt{constant*tbabs*(simplcutx*(diskbb + relxillionCp\_nk)+xillverCp)}   \\
		\hline
	\end{tabular}
\end{table*}

\subsection{Other setting and final results}

For the \textit{NuSTAR} data, we only set a cross-calibration constant between FPMA and FPMB. 
For the \textit{NuSTAR}+\textit{NICER} data, we add an {\tt edge} component to fit an absorption edge around 1.84 keV.
We also apply the {\tt xscat} model \citep{xscat_2016_ApJ_818_143S} for all data as the field of view of \textit{NICER} is comparable to the typical size of a dust scattering halo. We choose the MRN dust model for the scattering.
For the extraction regions, the radius is 50$^{\prime\prime}$ for the \textit{NuSTAR} data and 180$^{\prime\prime}$ for the \textit{NICER} data, in accordance with the configuration in \cite{Barillier_2023_ApJ_944_165B}.
Furthermore, we apply a cross-calibration model called {\tt crabber} \citep{crabcor_Steiner_2010ApJ} which is designed to standardize the calibration parameters between different detectors for the Crab Nebula (power-law normalization and index) via a multiplicative power-law correction.
All parameters in the \textit{NuSTAR} and \textit{NICER} spectra are linked except the cross-calibration parameters.

For both Model 1 and Model 2, we link $\Gamma$ and $kT_{e}$ between the power-law component ({\tt{simplcutx}} or {\tt nthcomp}) and the reflection component ({\tt{relxillionCp\_nk}} and {\tt xillverCp}). The reflection fraction $R_{\rm f}$ in {\tt relxillionCp\_nk} is set to $-1$, so the model returns only the reflection component. We also link the neutral column in {\tt xscat} to the value from  {\tt tbabs}.

We apply the AICc \citep{Burnham_2002_book}, which is the Akaike information criterion (AIC) \citep{Akaike_1974_ITAC_19} corrected for small sample sizes and provides a mean for selecting the best model among several model candidates. The model candidate with the lowest AICc is the best model. The formulas used to calculate AIC and AICc are
\begin{equation}
	{\rm AIC} = -2{\rm ln}\mathcal{L} + 2N_{p}
\end{equation}
and
\begin{equation}
	{\rm AICc} = {\rm AIC} + 2N_{p}(N_{p} + 1)/(N_{b} - N_{p} - 1),
\end{equation}
where $N_{p}$ is the number of free parameters, $N_{b}$ is the number of bins and $\mathcal{L}$ is the likelihood of the best-fit model. The values of likelihood can be written in terms of $\chi^2$ which is obtained from the model's best fit: $\chi^2 = -2 \ln \mathcal{L}$.
\par
\begin{figure*}
	\centering
	\caption{Model and residual plots for EXO~1846--031 with \textit{NuSTAR} data. In the left panel, the red lines represent the thermal components, the blue lines represent the power-law components, and orange lines represent the reflection components. In the right panel, the blue lines represent the Comptonized thermal spectra, and the orange lines represent the Comptonized reflection spectra. For the lines with same color, solid lines represent the complete spectra, dotted lines represent the components scattered by the corona, and dashed lines represent the components transmitted by the corona.}
	\label{NuSTAR_plot}
	\vspace{0.5cm}
	\includegraphics[width=0.95\linewidth]{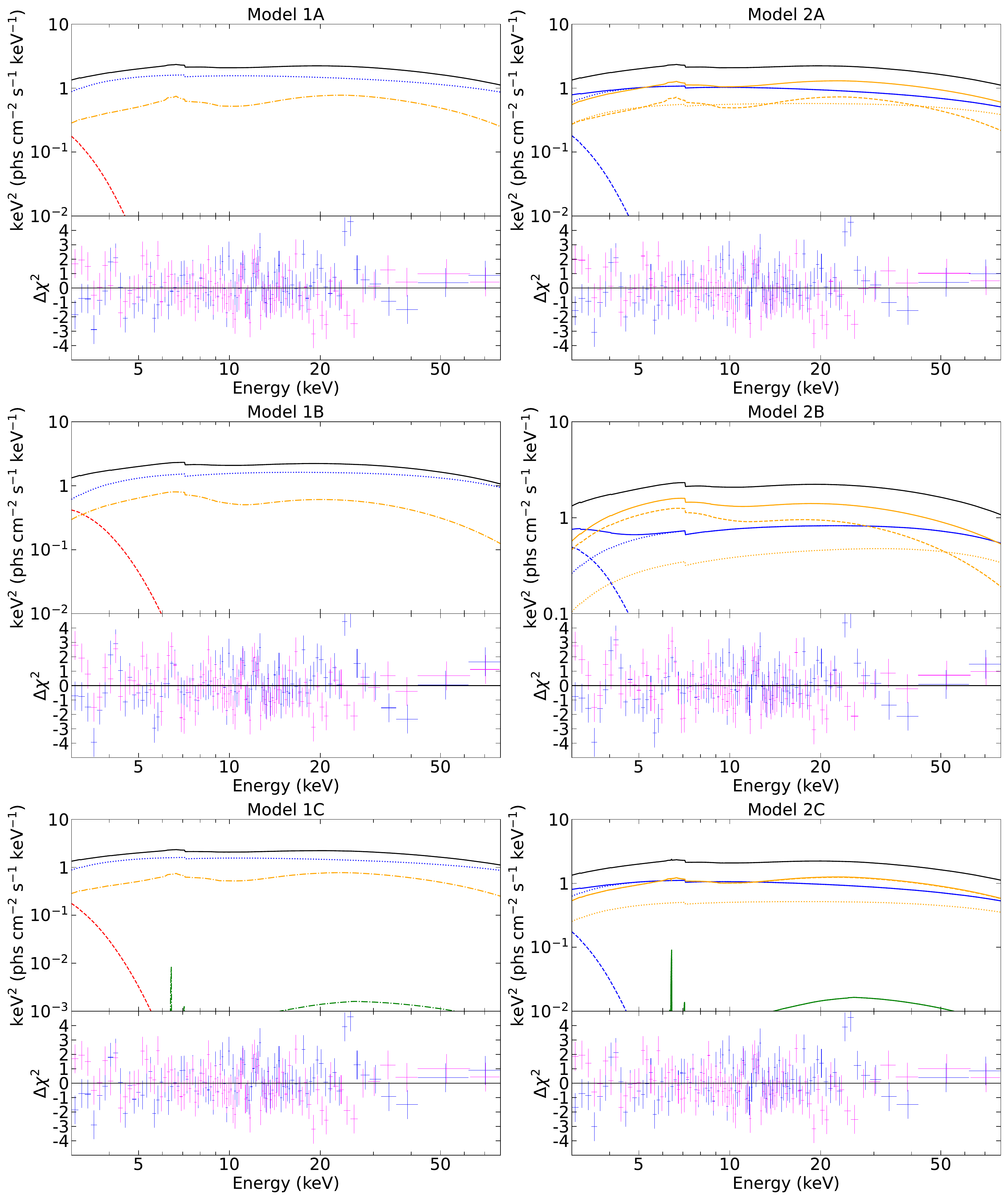}
	\vspace{0.5cm}
\end{figure*}

For the \textit{NuSTAR} data, the best-fit for all models are presented in Table~\ref{bestfit:EXO1846_NuSTAR} and the best-fit model plots and residual plots are presented in Fig.~\ref{NuSTAR_plot}.

\begin{table*}
	\centering
	\renewcommand\arraystretch{1.5}
	\caption{Best-fit values of all models with \textit{NuSTAR} data. $P$ means that we reach the lower or upper limit of the model. The reported uncertainties correspond to the 90\% confidence intervals.}
	\label{bestfit:EXO1846_NuSTAR}
	\vspace{0.3cm}
	{
		\begin{tabular}{lcccccc}
			\hline
			Model                                     & Model 1A                        & Model 1B                          & Model 1C                        & Model 2A                      & Model 2B                          & Model 2C                           \\
			\hline
			{\tt tbabs}                                                                                                                                                                                                                                                \\
			$N_{\rm H}$ [\(10^{22}\) cm\(^{-2}\)]     & $7.7_{-0.5}^{+0.5}$             & $9.7_{-0.6}^{+0.4}$               & $7.66_{-0.21}^{+0.5}$           & $7.7_{-0.6}^{+0.5}$           & $10.5_{-0.5}^{+0.6}$              & $7.7_{-0.4}^{+0.6}$                \\
			\hline
			{\tt diskbb}                                                                                                                                                                                                                                               \\
			$T_{\rm in}$                              & $0.33_{-0.04}^{+0.04}$          & $0.452_{-0.011}^{+0.011}$         & $0.33_{-0.06}^{+0.04}$          & $0.349_{-0.014}^{+0.019}$     & $0.459_{-0.004}^{+0.011}$         & $0.337_{-0.015}^{+0.012}$          \\
			Norm                        & $1.1_{-0.6}^{+2.4}\times10^{5}$ & $1.88_{-0.4}^{+0.23}\times10^{4}$ & $1.1_{-0.6}^{+2.3}\times10^{5}$ & $8.6_{-1.8}^{+6}\times10^{4}$ & $2.48_{-0.4}^{+0.23}\times10^{4}$ & $1.13_{-0.11}^{+0.25}\times10^{5}$ \\
			\hline
			{\tt simplcutx} \\
			$\Gamma$                                  &       &          &         & $2.139_{-0.014}^{+0.012}$     & $1.9345_{-0.0026}^{+0.03}$        & $2.14_{-0.02}^{+0.03}$             \\
			$kT_{\rm e}$ [keV]                        &            &             &             & $58.7_{-16.3}^{+9}$           & $31.6_{-1.6}^{+1.8}$              & $66_{-18.8}^{+30.0}$               \\
			$f_{\tt SC}$                              &                               &                                &                               & $0.215_{-0.026}^{+0.013}$     & $0.120_{-0.04}^{+0.024}$          & $0.196_{-0.03}^{+0.015}$           \\
			\hline
			{\tt nthcomp}                                                                                                                                                                                                                                              \\
			$\Gamma$                                  & $2.102_{-0.015}^{+0.028}$       & $1.986_{-0.011}^{+0.007}$         & $2.102_{-0.04}^{+0.024}$        &      &         &              \\
			$kT_{\rm e}$ [keV]                        & $61_{-17.8}^{+26.9}$            & $31.1_{-2.1}^{+4}$                & $61_{-20.7}^{+17.6}$            &           &               &               \\
			Norm                       & $1.70_{-0.13}^{+0.16}$          & $1.12_{-0.13}^{+0.03}$            & $1.70_{-0.14}^{+0.16}$          &                              &                                 &                                  \\
			\hline
			{\tt relxillionCp\_nk}                                                                                                                                                                                                                                     \\
			$q_{\rm in}$                              & $9.8_{-2.1}^{+P}$               & $10.0_{-0.3}^{+P}$                & $9.8_{-2.1}^{+P}$               & $9.5_{-1.4}^{+0.4}$           & $10.0_{-1.4}^{+P}$                & $9.5_{-1.7}^{+0.5}$                \\
			$q_{\rm out}$                             & $0.17_{-P}^{+0.9}$              & $3^{*}$                                 & $0.17_{-P}^{+1.1}$              & $0.0011_{-P}^{+1.1}$          & $3^{*}$                                 & $0.8_{-0.4}^{+0.3}$                \\
			$R_{\rm br}$ [$R_{\rm g}$]               & $6.6_{-1.4}^{+2.9}$             & $3.00_{-0.12}^{+0.29}$            & $6.6_{-1.5}^{+1.0}$             & $7.5_{-0.9}^{+1.2}$           & $3.18_{-0.17}^{+0.4}$             & $5.1_{-1.5}^{+1.7}$                \\
			$a_{*}$                                   & $0.9912_{-0.0028}^{+0.0019}$    & $0.973_{-0.016}^{+0.005}$         & $0.9912_{-0.0026}^{+0.002}$     & $0.9914_{-0.0022}^{+0.003}$   & $0.953_{-0.024}^{+0.023}$         & $0.9918_{-0.0027}^{+0.0022}$       \\
			$i$ [deg]                                 & $76.5_{-1.5}^{+1.2}$            & $67.7_{-2.7}^{+1.4}$              & $76.5_{-0.5}^{+1.3}$            & $76.4_{-0.6}^{+0.6}$          & $62.7_{-2.4}^{+7}$                & $76.8_{-0.6}^{+1.4}$               \\
			A$_{\rm Fe}$                              & $0.88_{-0.15}^{+0.18}$          & $2.20_{-0.5}^{+0.28}$             & $0.88_{-0.08}^{+0.21}$          & $0.91_{-0.17}^{+0.3}$         & $10.0_{-1.3}^{+P}$                & $0.88_{-0.06}^{+0.5}$              \\
			$\alpha_{\xi}$                            & $0.18_{-0.04}^{+0.05}$          & $0.001_{-P}^{+0.08}$              & $0.178_{-0.05}^{+0.019}$        & $0.176_{-0.028}^{+0.026}$     & $0.14_{-P}^{+0.16}$               & $0.18_{-0.05}^{+0.04}$             \\
			log${\xi}$ [erg~cm~s$^{-1}$] & $3.13_{-0.11}^{+0.18}$          & $3.60_{-0.08}^{+0.1}$             & $3.13_{-0.11}^{+0.12}$          & $3.10_{-0.04}^{+0.12}$        & $4.52_{-0.16}^{+0.04}$            & $3.10_{-0.07}^{+0.19}$             \\
			Norm                  & $0.0061_{-0.0009}^{+0.0011}$    & $0.0038_{-0.0003}^{+0.0003}$      & $0.0061_{-0.0006}^{+0.001}$     & $0.0083_{-0.0012}^{+0.0009}$  & $0.00526_{-0.00024}^{+0.0008}$    & $0.0083_{-0.0018}^{+0.0012}$       \\
			\hline
			{\tt xillverCp}                                                                                                                                                                                                                                            \\
			Norm                    &                               &                               & $0.0001_{-P}^{+0.004}$          &                         &                                 & $0.0012_{-P}^{+0.002}$             \\
			\hline
			$C_{\rm FPMB}$                            & $1.0152_{-0.0014}^{+0.0014}$    & $1.0152_{-0.0014}^{+0.0014}$      & $1.0152_{-0.0014}^{+0.0014}$    & $1.0152_{-0.0014}^{+0.0014}$  & $1.0152_{-0.0014}^{+0.0014}$      & $1.0152_{-0.0014}^{+0.0013}$       \\
			\hline
			$\chi^2$ /dof  & 633.03/513 & 695.0/514 & 633.0/512 & 632.85/513 & 701.7/514 & 632.7/512 \\
			  & =1.23398 & =1.35214 & =1.23633 & =1.23363 & =1.36518 & =1.23574 \\
			AICc                                      & 666.09                          & 725.94                            & 668.2                           & 665.91                        & 732.64                            & 667.9                              \\
			\hline
		\end{tabular}
	}
\end{table*}

\begin{figure*}
	\centering
	\caption{Model and residual plots for EXO~1846--031 with \textit{NuSTAR}+\textit{NICER} data. In the left panel, the red lines represent the thermal components, the blue lines represent the power-law components, and orange lines represent the reflection components. In the right panel, the blue lines represent the Comptonized thermal spectra and the orange lines represent the Comptonized reflection spectra. For the lines with same color, solid lines represent the complete spectra, dotted lines represent the components scattered by the corona, and dashed lines represent the components transmitted by the corona.}
	\label{NuSTARandNICER_plot}
	\vspace{0.3cm}
	\includegraphics[width=0.95\linewidth]{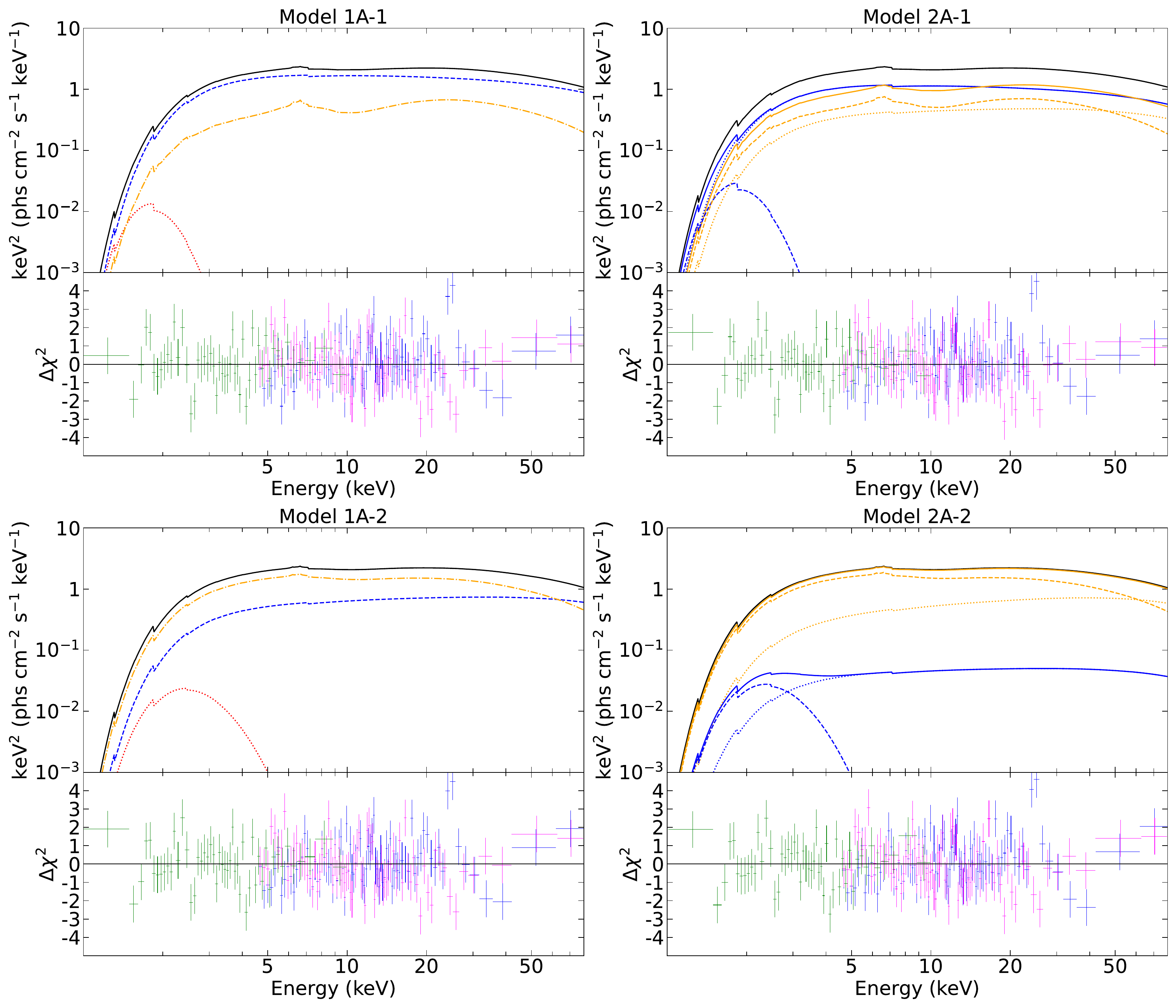}
\end{figure*}

As for the \textit{NuSTAR}+\textit{NICER} data, the best-fit for all models are presented in Table~\ref{bestfit:NuSTAR+NICER} and the best-fit model plots and residual plots are presented in Fig.~\ref{NuSTARandNICER_plot}. In addition, for all the plots of best-fit models with {\tt simplcutx}, we also plot the scattered and transmitted spectra separately.

\begin{table*}
	\centering
	\renewcommand\arraystretch{1.5}
	\caption{Best-fit values and unabsorbed fluxes of all models with \textit{NuSTAR}+\textit{NICER} data. All models have $q_{\rm out}$ free in the fit and there is no distant reflector (case A). Models 1A-1 and 2A-1 are the fits with the lowest values of $\chi^2$ (respectively without and with {\tt simplcutx}). Models 1A-2 and 2A-2 are local minima, as described in Section~\ref{soft x-ray data} (respectively without and with {\tt simplcutx}). $P$ means that we reach the lower or upper limit of the model. In the estimate of the Eddington ratio (last row), $D$ is the distance in kpc and $M$ is the mass in units of $M_{\odot}$. The estimate of the total flux is obtained by integrating the total spectrum of the best-fit model from 0.01 to 1000 keV. The reported uncertainties correspond to the 90\% confidence intervals.}
	\label{bestfit:NuSTAR+NICER}
	\vspace{0.3cm}
	\begin{tabular}{lcccc}
	\hline
	
	Model & Model 1A-1 & Model 2A-1 & Model 1A-2 & Model 2A-2 \\
	
	\hline
	
	{\tt edge}\\
	
	$\tau_{\tt max}$ & $0.095_{-0.022}^{+0.026}$ & $0.098_{-0.018}^{+0.015}$ & $0.122_{-0.015}^{+0.016}$ & $0.115_{-0.014}^{+0.024}$\\

	\hline

	{\tt tbabs}\\

	$N_{\rm H}$ [\(10^{22}\) cm\(^{-2}\)] & $6.33_{-0.1}^{+0.09}$ & $6.48_{-0.11}^{+0.11}$ & $5.60_{-0.05}^{+0.29}$ & $5.82_{-0.19}^{+0.07}$\\

	\hline

	{\tt xscat}\\

	$x$ & $1.00_{-0.05}^{+P}$ & $1.00_{-0.05}^{+P}$ & $0.90_{-0.12}^{+0.08}$ & $0.99_{-0.13}^{+P}$\\
	
	\hline

	{\tt diskbb}\\

	$T_{\rm in}$ & $0.192_{-0.01}^{+0.008}$ & $0.202_{-0.01}^{+0.006}$ & $0.42_{-0.06}^{+0.03}$ & $0.40_{-0.09}^{+0.04}$\\
	
	Norm & $6.2_{-2.7}^{+2.8}\times10^{5}$ & $7.59_{-0.04}^{+4}\times10^{5}$ & $1098_{-668}^{+2152}$ & $1875_{-1156}^{+2723}$\\

	\hline

	{\tt simplcutx}\\

	$\Gamma$ &  & $2.098_{-0.007}^{+0.017}$ &  & $1.907_{-0.007}^{+0.023}$\\

	$kT_{\rm e}$ [keV] & & $44.3_{-5}^{+1.4}$ &  & $36.1_{-1.8}^{+6}$\\

	$f_{\tt SC}$ &    & $0.229_{-0.028}^{+0.08}$ &    & $0.150_{-0.04}^{+0.021}$\\

	\hline

	{\tt nthcomp}\\
	
	$\Gamma$ & $2.052_{-0.004}^{+0.026}$ &  & $1.875_{-0.007}^{+0.007}$ & \\
	
	$kT_{\rm e}$ [keV] & $38.0_{-2.5}^{+1.6}$ &  & $41.4_{-1.8}^{+1.2}$ & \\
	
	Norm & $1.935_{-0.012}^{+0.018}$ &   & $0.376_{-0.015}^{+0.14}$ &  \\
	\hline

	{\tt relxillionCp\_nk}\\

	$q_{\rm in}$ & $10.0_{-1.4}^{+P}$ & $10.00_{-0.22}^{+P}$ & $9.8_{-0.9}^{+P}$ & $9.9_{-0.7}^{+P}$\\
	
	$q_{\rm out}$  & $0.46_{-0.12}^{+0.6}$ & $0.52_{-0.09}^{+0.15}$ & $1.44_{-0.26}^{+0.08}$ & $1.259_{-0.5}^{+0.028}$\\
	
	$R_{\rm br}$ [$R_{\rm g}$] & $5.59_{-0.26}^{+0.2}$ & $5.5_{-0.3}^{+0.8}$ & $4.45_{-0.29}^{+0.8}$ & $5.10_{-0.6}^{+0.13}$\\
	
	$a_{*}$ & $0.9912_{-0.0025}^{+0.0015}$ & $0.990_{-0.002}^{+0.004}$ & $0.9851_{-0.004}^{+0.0017}$ & $0.978_{-0.002}^{+0.007}$\\
	
	$i$ [deg] & $75.9_{-1.3}^{+0.7}$ & $75.1_{-1.3}^{+1.3}$ & $68.98_{-6}^{+0.27}$ & $68.4_{-0.4}^{+0.6}$\\
	
	A$_{\rm Fe}$ & $1.40_{-0.3}^{+0.19}$ & $0.867_{-0.016}^{+0.1}$ & $7.5_{-1.0}^{+0.3}$ & $6.60_{-1.8}^{+0.26}$\\
	
	$\alpha_{\xi}$ & $0.179_{-0.026}^{+0.04}$ & $0.204_{-0.012}^{+0.04}$ & $1.11_{-0.21}^{+0.5}$ & $1.03_{-0.3}^{+0.24}$\\
	
	log${\xi}$ [erg~cm~s$^{-1}$] & $3.15_{-0.04}^{+0.09}$ & $3.27_{-0.22}^{+0.11}$ & $4.70_{-0.03}^{+P}$ & $4.70_{-0.11}^{+P}$\\
	
	Norm & $0.00461_{-0.0002}^{+0.00011}$ & $0.0073_{-0.0008}^{+0.0007}$ & $0.0071_{-0.0011}^{+0.0009}$ & $0.00117_{-2.8e-05}^{+3.8e-05}$\\

	\hline
	
	$\rm dGamma_{\rm NICER}$ & $-0.172_{-0.016}^{+0.013}$ & $-0.175_{-0.019}^{+0.013}$ & $-0.169_{-0.015}^{+0.018}$ & $-0.137_{-0.017}^{+0.02}$\\
	
	$\rm crabberNorm_{\rm NICER}$ & $0.745_{-0.015}^{+0.019}$ & $0.741_{-0.04}^{+0.016}$ & $0.738_{-0.016}^{+0.05}$ & $0.792_{-0.027}^{+0.016}$\\
	
	$\rm crabberNorm_{\rm FPMB}$ & $1.0142_{-0.0015}^{+0.0015}$ & $1.0142_{-0.0015}^{+0.0015}$ & $1.0142_{-0.0015}^{+0.0015}$ & $1.0142_{-0.0008}^{+0.0015}$\\

	\hline
	
	$\chi^2$ /dof & 804.3/706 & 805.4/705 & 812.8/706 & 816.4/705 \\
	  & =1.13924 & =1.14241 & =1.15127 & =1.15801 \\
	
	AICc & 845.49 & 846.59 & 853.99 & 857.59\\
	
	\hline
     
	disk/total & 0.53 & 0.46 & 0.036 & 0.044 \\ 

	total flux [$10^{-8}\rm ergs/\rm cm^2/\rm s$] & 5.56 & 4.58 & 1.99 & 2.01 \\ 
        
	Eddington ratio (total) [$\frac{D^2}{M}$] & 0.053 & 0.044 & 0.019 & 0.019 \\
	
	\hline
	
	\end{tabular}
	
\end{table*}

\section{Discussion}
\label{discussion}

\subsection{Impact of reflection Comptonization}

In our study, we use \textit{NICER} and \textit{NuSTAR} data of EXO~1846--031 to investigate the impact of Comptonization of the reflection component on the measurements of the spectral parameters.
In Model 2, the Comptonized reflection spectrum is produced by {\tt simplcutx}*{\tt relxillionCp\_nk}, while there is no Comptonized reflection spectrum in Model 1.

\begin{figure*}
	\centering
	\caption{The 1-D $\chi^2$  contours for black hole spin (left) and inclination angle (right) for Model 1A and Model 2A with \textit{NuSTAR} data. The horizontal dashed line represents the $1\sigma$, $2\sigma$, $3\sigma$ confidence levels for a single parameter. }
	\label{steppar}
	\includegraphics[width=\linewidth]{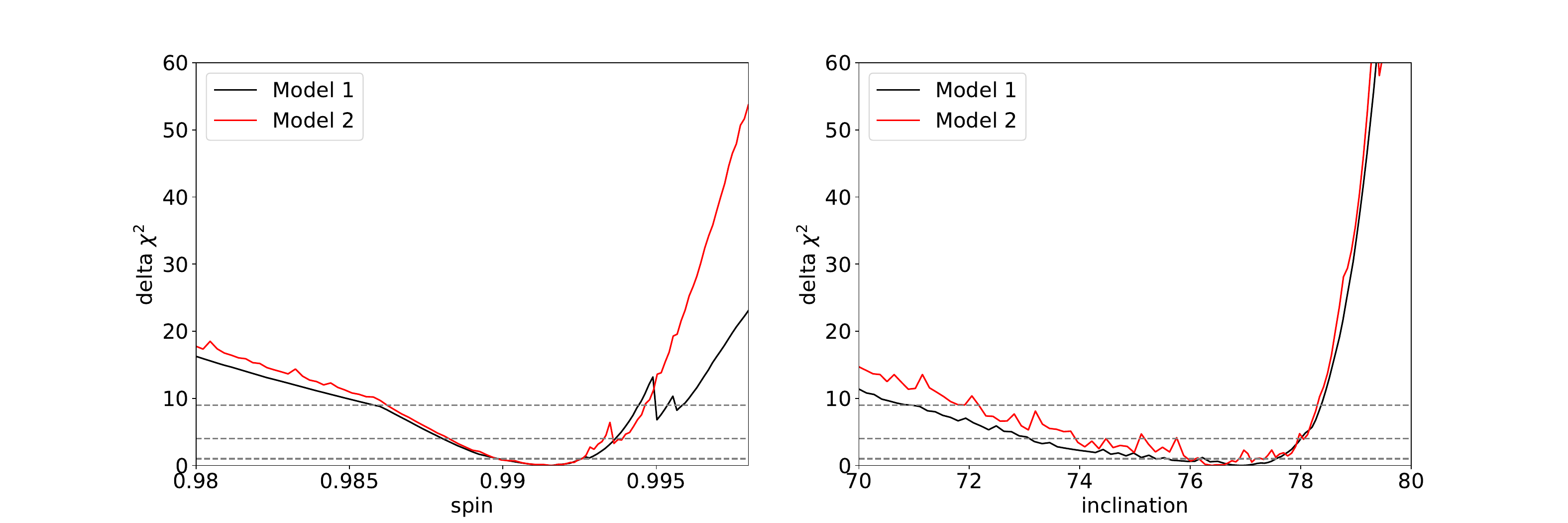}
	\vspace{0.5cm}
\end{figure*}

We do not see significant differences in the parameter estimation between Model 1 and Model 2 (Table~\ref{bestfit:EXO1846_NuSTAR} and Table~\ref{bestfit:NuSTAR+NICER}). The key-parameters, such as $a_{*}$ and $i$, are quite similar at $3\sigma$ confidence level (see Fig.~\ref{steppar}), indicating that the parameter measurements are not affected.
In addition, the reduced $\chi^2$ and AICc are also very close to each other.
Possibly, stronger differences in those key parameters would manifest in the case of larger $f_{\rm SC}$ (viz: $f_{\rm SC} > $ 0.2).
Other studies also suggest that the inclusion of reflection Comptonization does not significantly influence reflection spectrum fitting; see, e.g., Refs.~\cite{Dong_2022_MNRAS, Liu_2023_APJ, Liu_Qichun_2022_MNRAS}.

The scattered fraction $f_{\rm SC}$ in {\tt simplcutx} depends on the coronal geometry and optical depth. 
\citet{Liu_2023_APJ} found that the scattering fraction decreases during the hard-to-soft transition of GX~339--4 in the outburst in 2021 observed by \textit{Insight}-HXMT.
\citet{Wang_2022_ApJ} studied the reverberation lag of EXO~1846--031 to infer the disk-corona geometry and concluded that the corona may be the base of the jet and that the vertical extension of the jet changes during the hard-to-soft transition.
These studies support the idea of an evolving corona across state transition in X-ray binaries, one measure of which is the scattering fraction. In this case, although the model formalism with {\tt simplcutx} does not improve the fit statistically, in such a case looking for changes of $f_{\rm SC}$ with state may provide its own useful consistency check.

\subsection{Impact of soft X-ray data}
\label{soft x-ray data}

From the comparison of the results from the analyses of the \textit{NuSTAR} and \textit{NuSTAR}+\textit{NICER} data, we see the impact of the soft X-ray band in the analysis of the data.
The disk component peaks around 2.0-3.0 keV (see Fig.~\ref{NuSTARandNICER_plot}), but \textit{NuSTAR} lacks coverage below 3.0 keV. Consequently, the fit of the disk component with \textit{NuSTAR} data may not be as accurate as the fit with \textit{NuSTAR+NICER} data.
In terms of parameter estimation, while some parameters differ slightly (e.g. $N_{\rm H}$, $T_{\rm in}$), key parameters such as spin and inclination angle still do not change significantly.
It is tentatively suggested that the inclusion of soft X-ray data does not significantly impact the parameter estimations of the best-fit model.
However, by including \textit{NICER} soft X-ray data, we identify an additional $\chi^2$ minimum, not as deep, which exhibits a more appropriate flux of the disk component despite having a slightly higher AICc; see the columns of Models 1A-2 and 2A-2 in Table~\ref{bestfit:NuSTAR+NICER}. The main difference between 1A-1/2A-1 and 1A-2/2A-2 is in the flux of the thermal component of the disk. In the observations analyzed in this work, the source was in a hard-intermediate state. In 1A-1/2A-1, the Norm parameter of {\tt diskbb} exceeds $10^5$ and this component counts for about 50\% of the total unabsorbed flux. In \citet{Wang_2021}, the authors report the analysis of EXO~1846--031 in the soft state and find a value for the Norm parameter around 1500. These results are difficult to explain, as we should expect that the thermal component is stronger in the soft state. Moreover, from a very high value of the Norm parameter, we would infer that the inner radius is far from the radius of the innermost stable circular orbit (ISCO), see Eq.~(\ref{eq-normdiskbb}) below, which is the opposite of what we infer from the analysis of the reflection features. If we believe that the correct fits are 1A-2/2A-2, Norm is about 1000 and makes sense: the disk component is not stronger than in the soft state and the inner edge of the disk is around the ISCO.

The mass and distance of EXO~1846--031 have not been dynamically measured yet. Studies of spectral analysis suggest mass measurements ranging from 3.24 to 7.1-12.43 $M_{\odot}$ \citep{Strohmayer_2020_AAS_23515902S, Nath_2024_ApJ_960_5N}, and distance measurements ranging from 2.4 to 7 kpc \citep{Parmar_1993_A&A_279, Williams_2022_MNRAS_517_2801W}. 
Considering a distance of 7~kpc and a mass in the range 5 to 12~$M_{\odot}$, the Eddington ratio $L/L_{\rm Edd}$ is 21.6\%--51.7\% for Model 1A-1, 18\%--43\% for Model 2A-1, 7.7\%--13\% for Model 1A-2, and 7.8\%--13\% for Model 2A-2. 
Furthermore, the unabsorbed disk component is much fainter in Models 1A-2 and 2A-2 (refer to the ``disk/total'' row in Table~\ref{bestfit:NuSTAR+NICER}). 
This indicates that for certain values of mass and distance, the luminosity of Models 1A-1 and 2A-1 is too high, while the results of Models 1A-2 and 2A-2 are more appropriate.

From the norm of {\tt diskbb}, we can derive an approximate inner radius~\citep{diskbb_1984}:
\begin{equation}\label{eq-normdiskbb}
	{\rm Norm} = (r_{\rm in}/D_{10})^2 \; \cos\theta,
\end{equation}
where $D_{10}$ is the distance in the units of 10~kpc and $r_{\rm in}$ is the derived inner radius from the observed data. After factoring in some correction factors \citep{Kubota_1998_PASJ...50..667K}
\begin{equation}
	R_{\rm in} = \sqrt{\frac{3}{7}}\cdot \left(\frac{6}{7} \right)^3\cdot\kappa^2\cdot r_{\rm in},
\end{equation}
where $\kappa \sim 1.7$ is the ratio of color temperature to effective temperature \citep{Shimura_1995_ApJ...445..780S}, then we can obtain $R_{\rm in}$ which is the true inner radius.
For a Kerr black hole, $R_{\rm ISCO}$ can be derived from the spin \citep{Bardeen_1972_ApJ...178..347B}. 
Since we apply X-ray reflection spectroscopy to measure the spin, we should identify the $R_{\rm in}$ with $R_{\rm ISCO}$ \citep{X-ray_spectroscopy}.
Although the estimates of $R_{\rm in}$ and $R_{\rm ISCO}$ from observed data have significant uncertainty due to the lack of accurate measurements of mass and distance, $R_{\rm in}$ should not be much larger than $R_{\rm ISCO}$ given the reflection-fitting results.
$R_{\rm in}$ inferred from the results of Model 1A-1 and 2A-1 is about 9.7-70 $R_{\rm ISCO}$ (with the same uncertainty of mass and distance in estimation of Eddington ratio), while from the results of Model 1A-2 and 2A-2 is about 1.0-2.7 $R_{\rm ISCO}$.
Therefore, the best fits of Model 1A-2 and 2A-2 are somewhat favored.

\subsection{Impact of emissivity profile}

In the models with free $q_{\rm out}$, we get a flat emissivity profile at large radii with $q_{\rm out} \approx 0$. 
To validate these results, models with fixed $q_{\rm out}$ set to 3 were also examined, but they yielded significantly poorer best-fit results ($\Delta \rm AICc \sim 60$, as shown in Table~\ref{bestfit:EXO1846_NuSTAR} and Table~\ref{bestfit:NuSTAR+NICER}) compared to the best-fit with free $q_{\rm out}$ in both \textit{NuSTAR} data and \textit{NuSTAR}+\textit{NICER} data, with several parameters displaying notable differences.
Additionally, models incorporating an additional {\tt xillverCp} component were tested to explore the possibility of obtaining a larger value for $q_{\rm out}$. However, the norm of {\tt xillverCp} is consistent with zero.
Furthermore, in the results of the \textit{NuSTAR+NICER} data in Model 1A-2 and Model 2A-2, the index $q_{\rm out} \gtrsim 1$, which alleviates concerns about a runway emissivity for $q_{\rm out} \sim 0$ (as illustrated in Fig.~\ref{relxill_Rout_qout}), so we only show the results of emissivity profile ``A'' in Table~\ref{bestfit:NuSTAR+NICER}. 
Other studies also reported instances of $q_{\rm out} \approx 0$ in GS~1354--645 \citep{GS1354_Xu_2018_ApJ} and GRS~1915+105 \citep{GRS1915_Zhang_2019_ApJ}. 
\par


\section{Conclusions}
\label{conclusion}

We investigated the impact of the Comptonization of the reflection spectrum in the case of EXO~1846--031. We have drawn the following conclusions:
\par
1. Considering reflection Comptonization by using {\tt simplcutx} does not yield a significantly different best fit and it does not affect the values of key parameters such as black hole spin and inclination angle. In the case of coronal properties evolving during state transitions, the variation in $f_{\rm SC}$ potentially provides a useful consistency check. 
\par
2. The inclusion of soft X-ray data does not significantly affect the estimations of key parameters, but gives a new fit with more appropriate luminosity, inner radius of disk and emissivity profile. The fit of \textit{NuSTAR+NICER} data with a weak disk component and non-zero $q_{\rm out}$ is the most appropriate in this case. 
\par


\vspace{0.5cm}

{\bf Acknowledgments --}
This work was supported by the Natural Science Foundation of Shanghai, Grant No.~22ZR1403400, and the National Natural Science Foundation of China (NSFC), Grants No.~12250610185, No.~11973019, and No.~12261131497.


\bibliography{bibliography}

\end{document}